\documentclass[a4paper,11pt,amsmath,amssymb]{article}
\usepackage{pos}
\usepackage{bm}% bold math

\title{Equivariant Transformer is all you need}
\author*[a,b]{Akio Tomiya}
\author[c,d]{Yuki Nagai}

\affiliation[a]{%
Faculty of Technology and Science, International Professional University of Technology, 3-3-1, Umeda, Kita-ku, Osaka, 530-0001, Osaka, Japan
}%
\affiliation[b]{
RIKEN/BNL Research center, Brookhaven National Laboratory, Upton, 11973, NY, USA
}

\affiliation[c]{CCSE, Japan Atomic Energy Agency, 178-4-4, Wakashiba, Kashiwa, Chiba 277-0871, Japan}
\affiliation[d]{
Mathematical Science Team, RIKEN Center for Advanced Intelligence Project (AIP), 1-4-1 Nihonbashi, Chuo-ku, Tokyo 103-0027, Japan
}

\emailAdd{akio@yukawa.kyoto-u.ac.jp}
\emailAdd{nagai.yuki@jaea.go.jp}

\abstract{
Machine learning, deep learning, has been accelerating computational physics, which has been used to simulate systems on a lattice. Equivariance is essential to simulate a physical system because it imposes a strong induction bias for the probability distribution described by a machine learning model. 
This reduces the risk of erroneous extrapolation that deviates from data symmetries and physical laws.
However, imposing symmetry on the model sometimes occur a poor acceptance rate in self-learning Monte-Carlo (SLMC). On the other hand, Attention used in Transformers like GPT realizes a large model capacity. We introduce symmetry equivariant attention to SLMC. To evaluate our architecture, we apply it to our proposed new architecture on a spin-fermion model on a two-dimensional lattice. We find that it overcomes poor acceptance rates for linear models and observe the scaling law of the acceptance rate as in the large language models with Transformers.
}

\FullConference{The 40th International Symposium on Lattice Field Theory (Lattice 2023)\\
July 31st - August 4th, 2023\\
Fermi National Accelerator Laboratory\\}

\begin{document}
\maketitle

\section{Introduction}
Lattice QCD is essential for calculating quantum field expectations but struggles with the critical slowing down issue, reducing computational efficiency. Machine learning methods can efficiently handle this problem and work well with structured data like gauge configurations \cite{Cranmer:2023xbe,Nagai2021-pw}.

Transformer is a neural network originally for dealing with natural languages, but it has been applied to various data \cite{Jumper2021,dosovitskiy2021image,Tholke2022-bg,Batzner2022-ns}. The most crucial feature of Transformer is it can deal with global correlations, such as modifiers in natural language, which sometimes act from distant places.

In computational physics with machine learning, in particular equivariant neural networks enhance numerical calculations by capturing input data symmetries. This improves generalization on unfamiliar data and may boost computational efficiency. Data symmetry ensures alignment with physical laws like momentum conservation. Convolutional layers, for instance, exhibit equivariant properties to spatial translation.
This reduces the risk of erroneous extrapolation that deviates from data symmetries and physical laws \cite{horie2023equivariant}.

In this work, we develop an equivariant Transformer for physical systems. In quantum field theory, local action/Hamiltonians are typically considered. However, when fermions, described by Grassmann numbers, are integrated ahead of numerical simulations, the resulting effective action/Hamiltonian becomes non-local regarding bosons.
We develop a neural network architecture which is capable with global correlations from fermions and symmetric under sytem's symmetries.

In this proof-of-principle study, we employ the {\it double exchange model} (DE) in two spacial dimensions.
DE model is well established model in condensed matter physics and contains fermions and spatially fixed classical Heisenberg-spins.
The model Hamiltonian is invariant under global O(3) spin rotation. 
This model is similar to Yukawa system with fermions and three component scalars on the lattice in particle physics.
To see more detail of this study, refer \cite{nagai2023selflearning}.

\section{Concepts in Machine learning}

\subsection{Self-learning Monte-Carlo}
We review concepts in machine learning to introduce our numerical calculation.
The Self-learning Monte-Carlo (SLMC) is an exact Markov chain Monte-Carlo (MCMC) algorithm with an effective model \cite{Liu_2017}.
%, which is consisted by two parts (Fig. \ref{fig:SLMC}).
In MCMC for a spin system, a spin configuration ${ {\bm S} }$ is distributed with a probability distribution $W({ {\bm S} })$. 
Samples from desired distribution are obtained after many steps.
The detailed balance condition is a sufficient condition for convergence of MCMC, which is
%\begin{align}
$W(\{ {\bm S} \} ) T(\{ {\bm S}' \}| \{ {\bm S} \}) = W(\{ {\bm S}' \} ) T(\{ {\bm S} \}| \{ {\bm S}' \})$,
%\end{align}
where $T(\{ {\bm S}' \}| \{ {\bm S} \}) $ is the transition probability from a configuration $\{ {\bm S} \}$ to another configuration $\{ {\bm S}' \}$.
If a probabilistic process described $T(\{ {\bm S}' \}| \{ {\bm S} \})$ with this condition, the obtained configurations distributed according to $W(\{ {\bm S} \} ) $.

In general Metropolis-Hastings (MH) algorithm, the transition probability is factorised in two sub-steps,
%\begin{align}
$T(\{ {\bm S}' \}| \{ {\bm S} \}) = g( 
\{ {\bm S}' \} |  
\{ {\bm S} \}) 
A( \{ {\bm S}' \}, \{ {\bm S} \})$,
%\end{align}
where the proposal distribution $g( \{ {\bm S} \} '|  \{ {\bm S} \})$ is the conditional probability of proposing a configuration $\{ {\bm S}' \}$ when a configuration $\{ {\bm S} \}$ is given, and the acceptance ratio $A( \{ {\bm S}' \}, \{ {\bm S} \})$ is the probability to accept the proposed configuration $A( \{ {\bm S}' \}, \{ {\bm S}' \})$. 
The Markov chain that has the desired distribution $W(\{ {\bm S} \} ) $ is obtained when the acceptance ratio is given as 
\begin{align}
A( \{ {\bm S}' \}, \{ {\bm S} \}) = {\rm min} \left( 1, \frac{W(\{ {\bm S}' \} ) }{W(\{ {\bm S} \} ) } \frac{g( \{ {\bm S} \} |  \{ {\bm S}' \})}{g( \{ {\bm S}' \} |  \{ {\bm S} \})} \right). \label{eq:A}
\end{align}
This is the general argument about the MH algorithm. 
The Metropolis test is a reduced version,
%\begin{align}
$
A( \{ {\bm S}' \}, \{ {\bm S} \}) = {\rm min} \left( 1, {W(\{ {\bm S}' \} ) }/{W(\{ {\bm S} \} ) }  \right) %, \label{eq:Met}
%\end{align}
$
and this is obtained assuming the reversibility for $g( \{ {\bm S}' \} |  \{ {\bm S} \})=g( \{ {\bm S} \} |  \{ {\bm S}' \})$.

\subsubsection{Inner Markov chain in SLMC}
SLMC is a nested MCMC. The simplest update is the local update, where a single site in the configuration is randomly selected and its spin orientation is changed.
We perform the Metropolis accept/reject procedure following to the single-site-update with $W_{\rm eff}(\{ {\bm S} \} ) $, which is the Boltzmann weight with an effective model.
Since the Metropolis test with the single cite update satisfies the detailed balance condition, it will converge into $W_{\rm eff}(\{ {\bm S} \} ) $.
In SLMC, the effective model $W_{\rm eff}(\{ {\bm S} \} ) $ contains trainable parameters.

\subsubsection{Outer Markov chain in SLMC}
\begin{figure}[t]
\begin{center}
\includegraphics[width=0.65\linewidth]{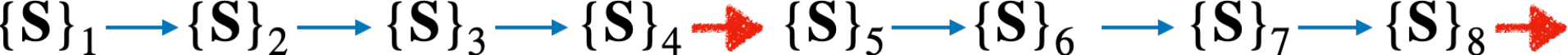}
\end{center}
\caption{
Schematic figure of SLMC.
Blue thin arrows the Metropolis chain in the inner chain with $W_{\rm eff}$.
Red bold arrows indicate the MH test in the outer chain with $W_{\rm eff}$ and $W$ as Eq. \eqref{eq:SLMC_acc}.
\label{fig:SLMC}
}
\end{figure}

In SLMC, the inner Markov chain using an effective model proposes processes for the outer chain. A correction step ensures the target system's distribution. In the general MH algorithm, the inner Markov updates represent the $g(\cdot|\cdot')$ process in the outer chain, with the acceptance ratio designed to offset $W_{\rm eff}$ and distribute as $W(\cdot)$.
The acceptance ratio in the SLMC is given as 
\begin{align}
    A( \{ {\bm S}' \}, \{ {\bm S} \}) = {\rm min} \left( 1, \frac{W(\{ {\bm S}' \} ) }{W(\{ {\bm S} \} ) } 
    \frac{W_{\rm eff}(\{ {\bm S} \} ) }{W_{\rm eff}(\{ {\bm S}' \} ) } 
    \label{eq:SLMC_acc}
\right),
\end{align}
where $W(\{ {\bm S}\} )$ is the probability weight for the target system.
We remark that, the second factor in the second column in $\min(1,\cdot)$ is up-side-down of the weights for the inner chain.
%If we can design the proposal (effective) Markov chain whose probability is equal to that of a Markov chain with the target Hamiltonian, the proposed spin configuration $\{ {\bm S}' \}$ is always accepted in the outer Metroplis-Hastings test. 
In summary, SLMC process can be express as Fig. \ref{fig:SLMC}. 

\subsection{Equivariance}

The concept of equivariance plays a pivotal role in machine learning and physics, offering solutions to issues like model overfitting and the preservation of physical laws. Its importance stems from its capacity to embed symmetries directly into neural networks, ensuring that the learned model respects certain symmetries of the data. 
%
%Mathematically, equivariance can be defined as follows. Given a set of data \( S \) and a transformation \( T \) that maps an element \( x \in S \) to \( x' = T(x) \), a function \( f: S \to S \) is said to be equivariant if it satisfies \( f(T(x)) = T(f(x)) \).
%
In neural networks, equivariance is usually achieved through weight sharing, reducing the number of irrelevant parameters. This is critical as an excess of parameters often leads to model overfitting, undermining the model's ability to generalize to new data.

In physics, ensuring numerical calculations align with physical laws can be achieved via equivariant neural networks. While penalties in the loss function were previously used to impose physical laws \cite{karniadakis2021physics}, this method is not always reliable. A more dependable approach embeds these symmetries directly into the neural network architecture \cite{horie2023equivariant}.

\section{Lattice setup}
In this study, we employ semi-classical {\it double exchange} (DE) model in two dimension for a testbed \cite{Barros2013-op,Liu2017-wl,Stratis2022-zr}. It is a semi-classical system and  has electrons and classical Heisenberg spins,
\begin{align}
H &= -t \sum_{\alpha,\langle i,j \rangle} (\hat{c}^{\dagger}_{i\alpha} \hat{c}_{j\alpha} + {\rm h.c.})  
+\frac{J}{2} \sum_{i} {\bm S}_i \cdot \hat{\bm \sigma}_i
\label{eq:exact_model},
\end{align} 
where ${\bm S}_i$ is the classical Heisenberg spins on the $i$-th site, namely it is normalized O(3) scalar field.
$\hat{c}_{i\alpha}^{\dagger}$  is the fermionic creation operator at the $i$-th site for fermion with spin $\alpha \in \{ \uparrow,\downarrow\}$. A symbol $\langle i,j \rangle$ indicate the pairs of nearest neighbors. The interaction term with Pauli matrices are defined as $[\hat{\bm \sigma}_i]_{\gamma} \equiv \hat{c}_{i\alpha}^{\dagger} \sigma_{\alpha \beta}^{\gamma} \hat{c}_{i\beta}$ ($\gamma = x,y,z$).
$J$ is the interaction strength between the classical spins and the electrons and
we consider the hopping constant $t$ as the unit of energy. 
We adopt the periodic boundary condition on $N_x \times N_y$ site system. The total number of the sites is $N \equiv N_x N_y$.
%We omit the chemical potential term for simplicity and this corresponds to the half-filling condition.
The Hamiltonian has $O(3)$ rotational symmetry in the spin sector and discrete translational invariance.% on the lattice. 
We want to calculate statistical expectation value with $\exp[-\beta H]$ for \eqref{eq:exact_model}.

\subsection{Equivariant Transformer for physical system}
Here we introduce an equivariant Transformer for a physical spin system.
Input of our Transformer is a spin configuration ${\bm S}$ and output is modified spin configuration.
In our formalism, query, key, and value are $N \times 3$ matrices and they are defined as 
\begin{align}
{\bm S}^{\rm Q} \equiv \hat{W}^{\rm Q} {\bm S}, \:
{\bm S}^{\rm K} \equiv \hat{W}^{\rm K} {\bm S}, \: 
{\bm S}^{\rm V} \equiv \hat{W}^{\rm V} {\bm S},
\end{align}
respectively. 
$S_{i \mu}^{\alpha} \equiv \sum_{\langle i,j \rangle_k} W^{\alpha}_k  S_{j \mu}$,
${W}^{\alpha}_k \in \mathbb{R}$ ($\alpha=$ Q, K, V and $k=0,1,2,\cdots, \tilde{N}$). A symbol $\langle i,j \rangle_k$ picks up $k$-th nearest neighbors for sites $i,j$. 
%See \cite{nagai2023selflearning} for detailed definition of $\hat{W}^{\alpha}$.
This procedure can be regarded as a block spin transformation with $\tilde{N}+1$ free parameters. In this work, we take $\tilde{N}=6$.
% In our system with classical spin field, 

Consequently, we introduce the following (equivariant) self-attention block as,
\begin{align}
{\rm SelfAttention}^{\rm spin}({\bm S}) = \check{M}{\bm S}^{\rm V},
\label{eq:self_attention}
\end{align}
where $\check{M}$ is a $N\times N$ matrix.
%\begin{align}
%\check{M}\equiv \sigma \left( {\bm S}^{{\rm Q}} {\bm S}^{{\rm K},\top} \big/{\sqrt{3}} \right), \label{eq:Mmatrix}
%\end{align}
This $\check{M}$ is defined by
%\begin{align}
$\left[\check{M} \right]_{ij} = \sigma \left(  \sum_{\mu=1}^3 S_{i\mu}^{\rm Q} S_{j\mu}^{\rm K} \big/ {\sqrt{3}}  \right)$,
%\end{align}
where $i, j$ are indices for spatial position on the lattice.
$\sigma(\cdot)$ is a nonlinear activation function.
Intuitively, the argument $\sum_\mu S_{i\mu}^{\rm Q} S_{j\mu}^{\rm K}$ is a set of 2 point correlation functions of the blocked spin field from a point $j$ to $i$ and this is invariant under O(3) rotations since rotation matrices cancel out.
In this study, we take an activation function $\sigma(\cdot)$ as ReLU function.
%Mathematically, the matrix $\check{M}$ is a functional of the Gram matrix with respect to ${\bm S}$. 
%The local operators $\hat{W}^{\rm Q},\hat{W}^{\rm K},\hat{W}^{\rm V}$ have trainable parameters. 
%It should be noted that the number of the trainable parameters in the local operator is usually fewer than a dozen. 
%For example, if one considers $k$-th neighbors, the number of the trainable parameters in one local operator is only $k+1$. 
%This operation corresponds to the weight sharing in machine learning.
%This definition is essential for the equivariance.

%We use the ReLU function for the activation function defined as 
%\begin{align}
%    {\rm ReLU}(x) = \left\{
%\begin{array}{ll}
%0 & x < 0, \\
%x & {\rm otherwise}
%\end{array}
%\right. .
%\end{align}

\begin{figure}[t]
\begin{center}
\includegraphics[scale=0.25]{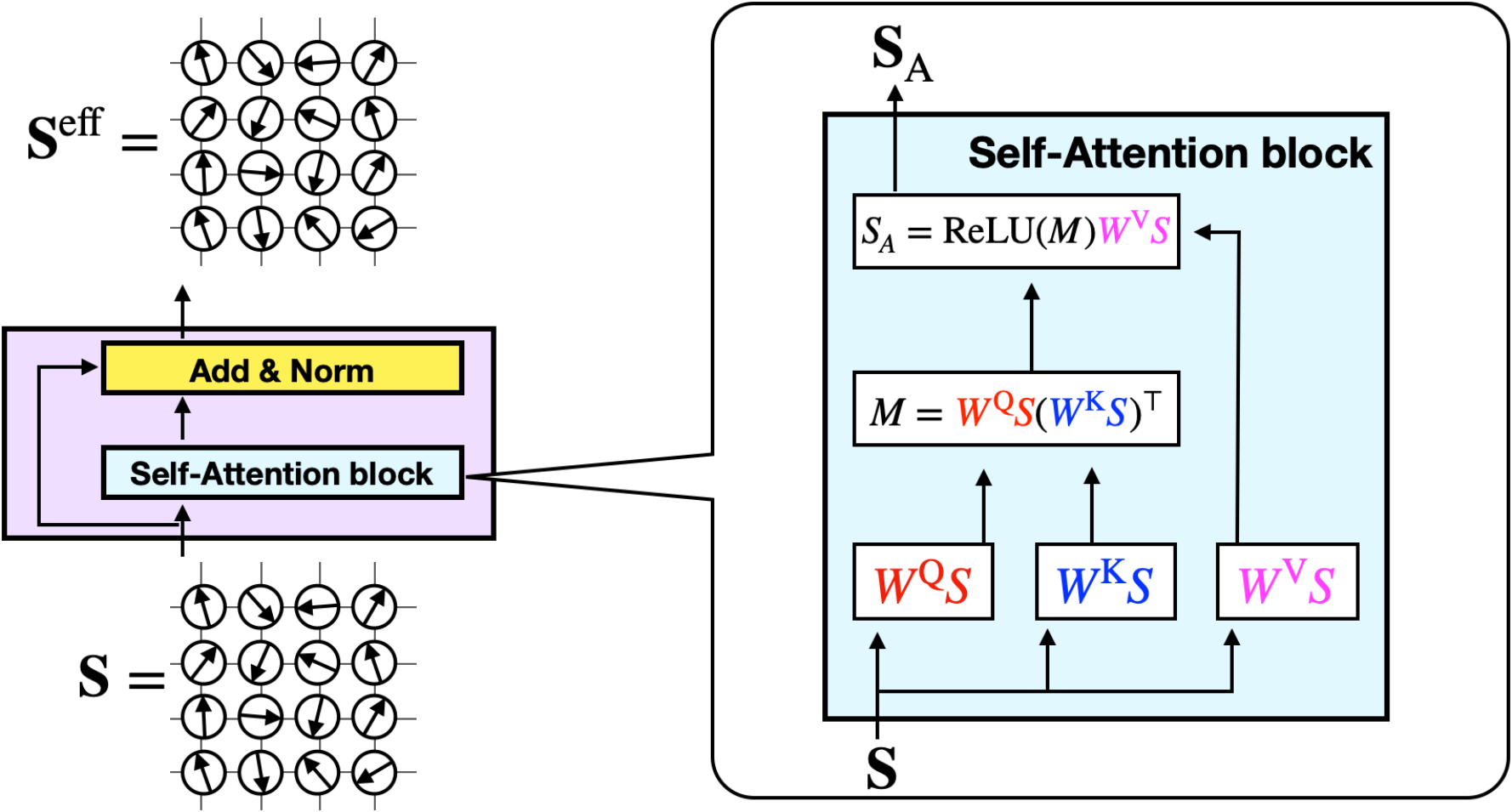}
\end{center}
  \caption{
(\textit{Left})
Effective spin construction using the Transformer with an Attention block. Yellow is defined by Eq. \eqref{eq:normalization_in_transformer}; purple is the attention block. (\textit{Right}) Blue represents the attention block (see main text).
   \label{fig:attention}}
\end{figure}

We construct the effective spin field that %consists of the neural networks 
with multiple attention layers. 
Our neural neural network architecture is defined with a residual connection, 
\begin{align}
    {\bm S}^{(l)} &\equiv {\cal N}\left({\bm S}^{(l-1)} + {\rm SelfAttention}_{{\bm \theta}^{(l)}}^{\rm spin}({\bm S}^{(l-1)}) \right), \label{eq:normalization_in_transformer} 
    %\\
    %\label{eq:transformer}
\end{align}
and
${\bm S}^{(0)} \equiv  {\bm S}$
and ${\bm S}^{\rm eff} \equiv  {\bm S}^{(L)}$.
$l$ is an index for layers and $l=1,2,\cdots, L$.
%$ {\rm SelfAttention}_{{\bm \theta}^{(l)}}^{\rm spin}({\bm S}^{(l-1)})$ is a neural network mapping in spin configurations space. %, which will be explained following.
${\bm \theta}^{(l)}$ represents a set of network trainable parameters in $l$-th layer. 
${\cal N}({\bm S})$  normalizes the spin vector on each lattice site,
$
{\cal N}({\bm S}_i) = {{\bm S}_i}/{\|{\bm S}_i\|}.
$
We call this network architecture the equivariant Transformer, which is schematically visualized in Fig.~\ref{fig:attention}.
We remark that if all weights $W^{\alpha}_k$ are 0 in $l$-th block, the self-attention block (indicated by a purple block in Fig.~\ref{fig:attention}) of $l$ works as an identity operation and it does nothing since the second term in the argument in \eqref{eq:normalization_in_transformer} is zero (See \cite{nagai2023selflearning} for details).
%We note that the effective spin is equivariant under the translational operation 

The long-range correlation in the DE model with SLMC is partially considered in the linear effective model in the literature \cite{Liu2017-wl,Kohshiro2021-ea}.
In this work, we replace a bare spin operator ${\bm S}_i$ by the effective spin operator from the Transformer as,
\begin{align}
H_{\rm eff}[{\bm S}] = - \sum_{\langle i,j \rangle_n} J_n^{\rm eff} {\bm S}_i^{\rm eff} \cdot {\bm S}_j^{\rm eff} + E_0, \label{eq:linear}
\end{align}
where ${\bm S}_i^{\rm eff}$ is an output the Transformer at site $i$. A symbol $\langle i,j \rangle_n$ represents the $n$-th nearest neighbor pair.
The effective spin ${\bm S}^{\rm eff}$ is a function of ${\bm S}$ and in total, $H_{\rm eff}$ is a function of ${\bm S}$.
This effective Hamiltonian contains a number of parameters in the Transformer.
In SLMC, $J_n^{\rm eff}$, $E_0$ and parameters in $ {\bm S}_i^{\rm eff}$ is determined by using AdamW and to increase acceptance ratio.

In our calculation, we apply the MH test \eqref{eq:SLMC_acc} using the effective model \eqref{eq:linear} and the Hamiltonian \eqref{eq:exact_model} in a form where the fermions is traced out by the exact diagonalization.

\section{Results}
\begin{figure}[t]
\begin{center}
\includegraphics[width=0.4\linewidth]{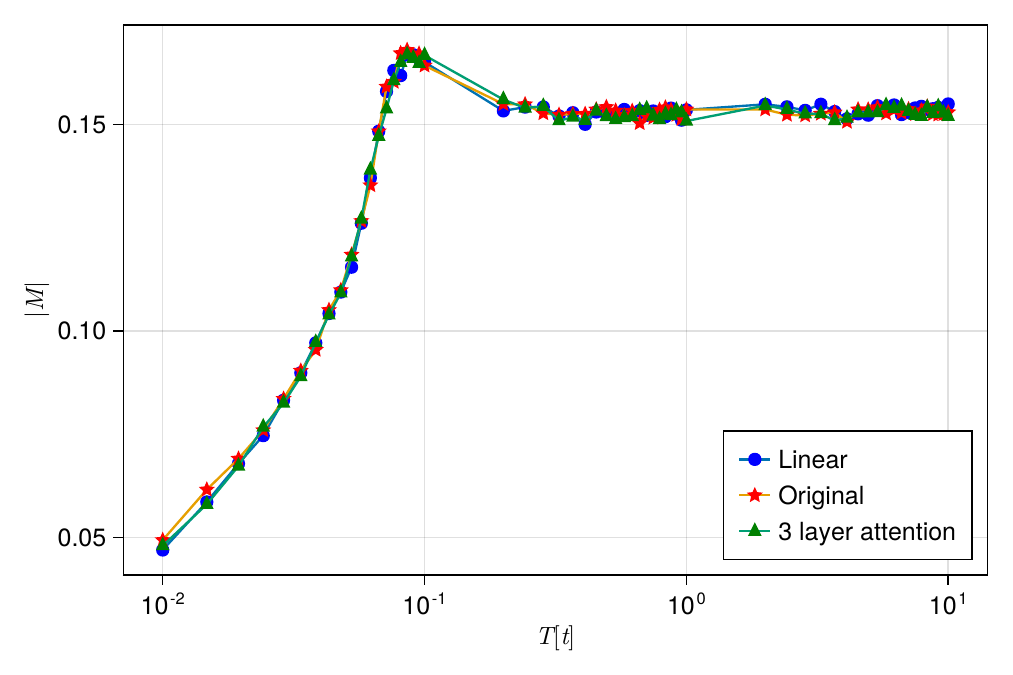}
\hspace{10mm}
\includegraphics[width=0.4\linewidth]{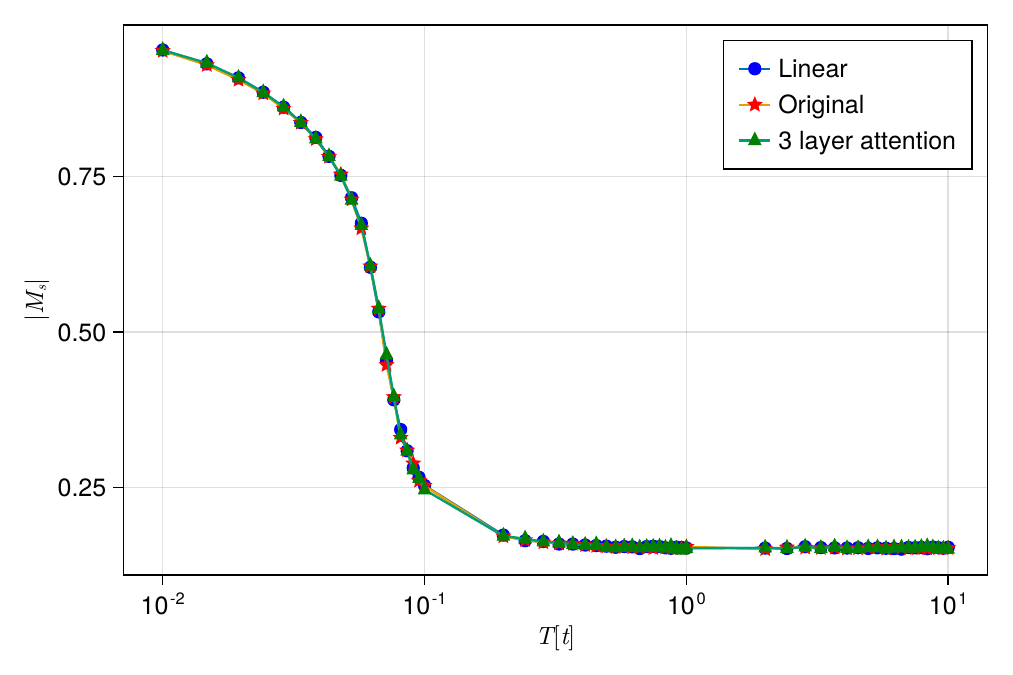}
\end{center}
  \caption{Magnetization (left panel) and staggered magnetization (right panel) for $6 \times 6$ lattice sites. 
For each temperature, we generate $2 \times 10^5$ samples using exact diagonalization (red stars), $5000$ samples using SLMC with the linear model (green triangles) and the effective model with attention blocks (blue circles). 
\label{fig:SLMCresults}}
\end{figure}

Here we show our results.
First, we show results for physical observables.
Our results show that it is consistent with exact results.
As depicted in Fig.~\ref{fig:SLMCresults}, the Self-Learning Monte Carlo (SLMC) method with effective models accurately replicates results from the original theory, exhibiting anti-ferromagnetic order at lower temperatures. 

\begin{figure}[th]
\begin{center}
\includegraphics[width=0.4\linewidth]{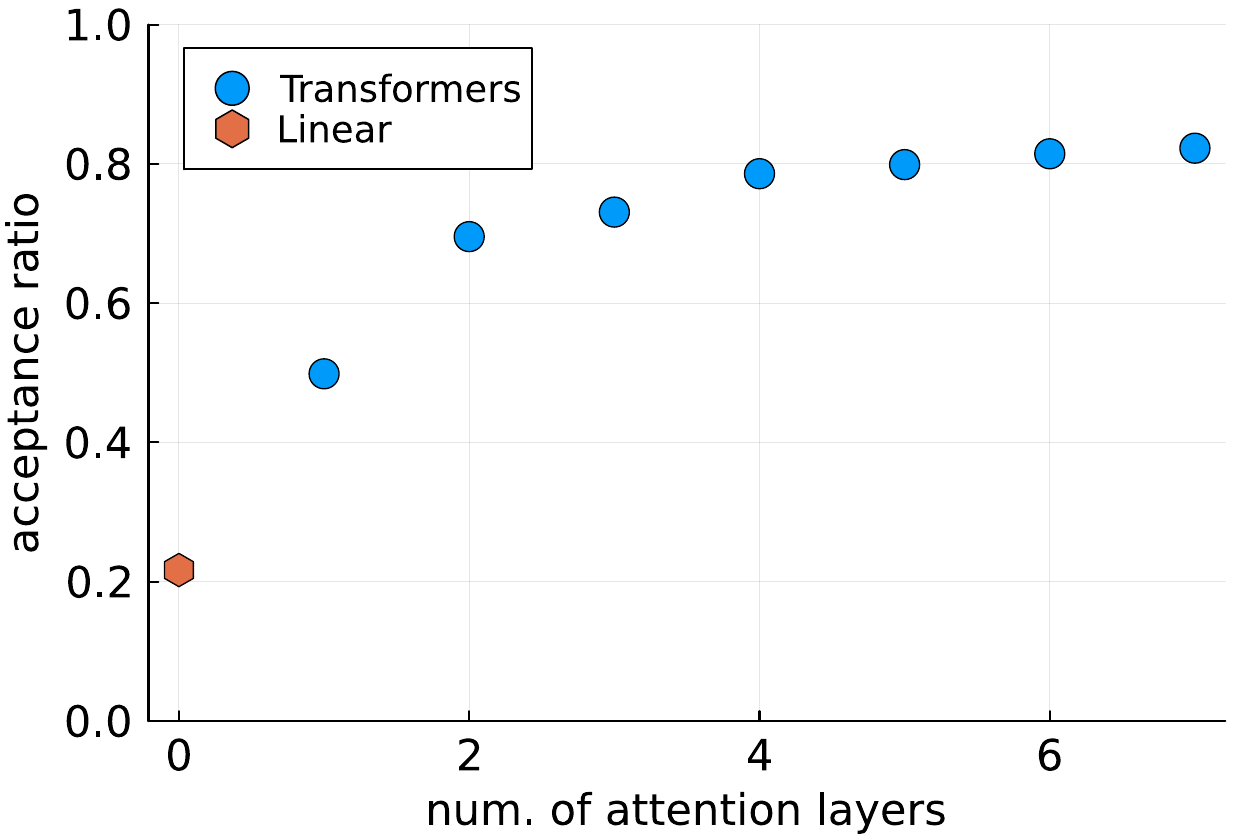}
%\end{center}
\hspace{10mm}
%\begin{center}
\includegraphics[width=0.4\linewidth]{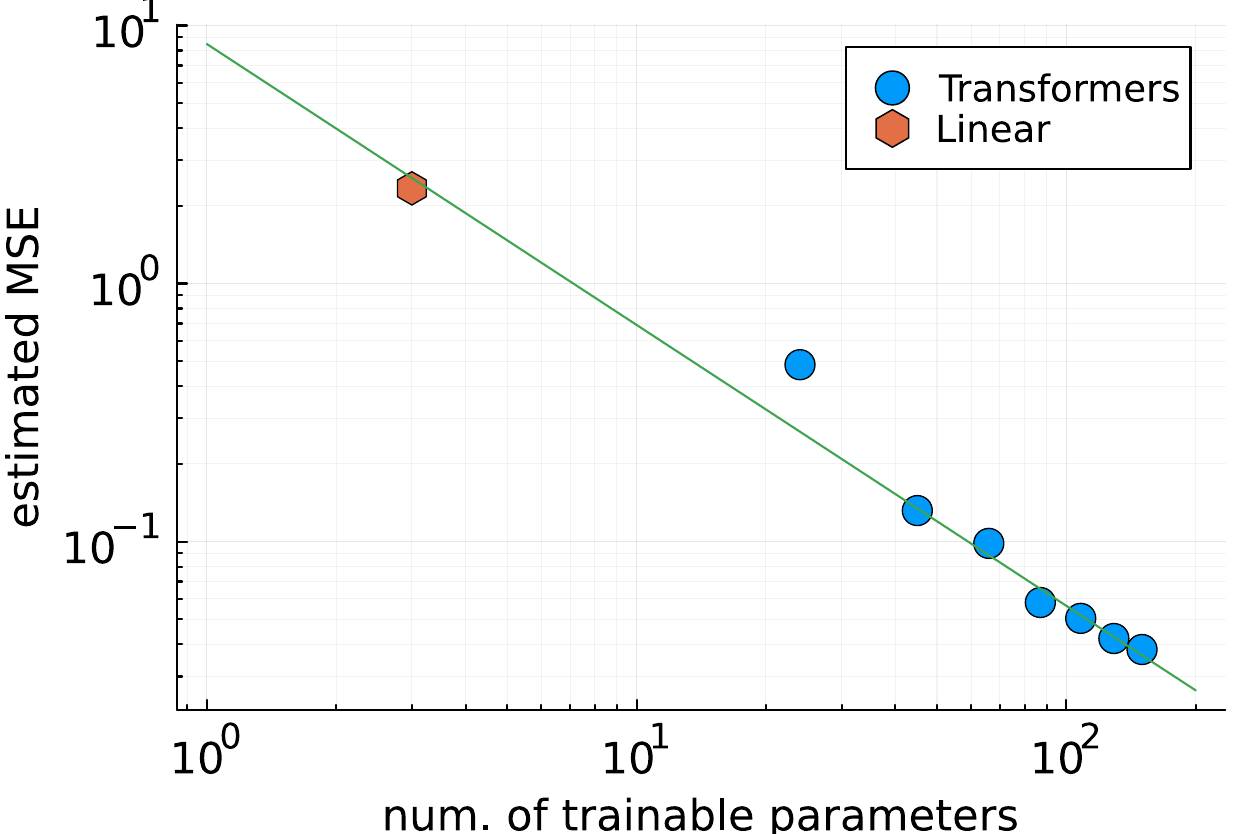}
\end{center}
  \caption{
({\it Left}) Acceptance ratio in SLMC: blue square for the linear model and red circles for attention models with $L=1,2,\cdots,6$. ({\it Right}) MSE vs. trainable parameters: blue square represents the linear model; red circles for models with $L=1,2,\cdots,6$. Fitting excludes the linear model and $L=1$.
\label{fig:scaling}\label{fig:layerdep}
}
\end{figure}

The acceptance rate with the number of layers is shown in Fig.~\ref{fig:layerdep}, left panel. Using effective models trained at \( T = 0.05t \) on a \( 6 \times 6 \) lattice, we set \( N_{\rm MC}^{\rm original} = 3 \times 10^4 \) and \( N_{\rm MC}^{\rm eff} = 100 \). The linear model SLMC has an acceptance ratio of 21\% due to the omission of long-range spin-spin interaction. As observed, the acceptance ratio improves with an increase in Attention layers.

Finally, we show results for the scaling law of the loss function in Fig.~\ref{fig:layerdep} (right panel).
The value of loss function is estimated from acceptance \cite{Shen2018-fl}.
It is known that, large language models with Transformer show a power-type scaling law; Model performance increases in depending on the size of the input data and the number of parameters in the model \cite{kaplan2020scaling}.
We find that, our model with equivariant Transformer shows the scaling law.
There are no direct relation from our model to large language models and the origin of the scaling law has to be studied in another work.

%\begin{figure}[t]
%\end{figure}

\section*{Acknowledgments}
The work of A.T. was partially by JSPS  KAKENHI Grant Numbers 20K14479, 22H05112, and 22H05111.
Y.N. was partially supported by JSPS KAKENHI Grant Numbers 22K12052, 22K03539, 22H05111 and 22H05114. 
The calculations were partially performed using the supercomputing system HPE SGI8600 at the Japan Atomic Energy Agency.
This work was partially supported by MEXT as ``Program for Promoting Researches on the Supercomputer Fugaku'' (Grant Number JPMXP1020230411, JPMXP1020230409).

%\begin{thebibliography}{99}
%\bibitem{...}
%\end{thebibliography}
\bibliographystyle{unsrt}
\bibliography{ref}% Produces the bibliography via BibTeX.
%\addbibresource{ref} 

\end{document}